\documentclass[aps,prc,groupedaddress,reprint,showpacs,floatfix]{revtex4-2}
\usepackage{graphicx}
\usepackage{xcolor}
\begin{document}
\title{Cross talk between experimental data and simple validation of shell closure in pre-actinides}
\author{Punit Dubey} \affiliation{Department of Physics, Banaras Hindu University, Varanasi-221005, India}
\author{Mahima Upadhyay} \affiliation{Department of Physics, Banaras Hindu University, Varanasi-221005, India}
\author{Mahesh Choudhary} \affiliation{Department of Physics, Banaras Hindu University, Varanasi-221005, India}
\author{Namrata Singh} \affiliation{Department of Physics, Banaras Hindu University, Varanasi-221005, India}
\author{Sriya Paul} \affiliation{Department of Physics, Banaras Hindu University, Varanasi-221005, India}
\author{Shweta Singh} \affiliation{Department of Physics, Banaras Hindu University, Varanasi-221005, India}
\author{N. Saneesh} \affiliation{Inter-University Accelerator Centre, Aruna Asaf Ali Marg, New Delhi-110067, India}
\author{Mohit Kumar} \affiliation{Inter-University Accelerator Centre, Aruna Asaf Ali Marg, New Delhi-110067, India}
\author{Rishabh Prajapati} \affiliation{Inter-University Accelerator Centre, Aruna Asaf Ali Marg, New Delhi-110067, India}
\author{K. S. Golda} \affiliation{Inter-University Accelerator Centre, Aruna Asaf Ali Marg, New Delhi-110067, India}
\author{Akhil Jhingan} \affiliation{Inter-University Accelerator Centre, Aruna Asaf Ali Marg, New Delhi-110067, India}
\author{P. Sugathan} \affiliation{Inter-University Accelerator Centre, Aruna Asaf Ali Marg, New Delhi-110067, India}
\author{Jhilam Sadhukhan} \affiliation{Variable Energy Cyclotron Centre, Kolkata-700064, India}
\author{Raghav Aggarwal} \affiliation{Department of Physics, Panjab University, Chandigarh-160014, India}
\author{Kiran} \affiliation{Department of Physics, Panjab University, Chandigarh-160014, India}
\author{Ajay Kumar}\email{ajaytyagi@bhu.ac.in}
\affiliation{Department of Physics, Banaras Hindu University, Varanasi-221005, India}
\date{\today}
%-------------------------------
\begin{abstract}
Two back-to-back experiments, $^{28}$Si + $^{178}$Hf and $^{28}$Si + $^{186}$W, were intentionally conducted to validate the role of shell closure in pre-actinides by studying neutron multiplicity in compound nucleus (CN) $^{206}$Rn and $^{214}$Ra. In the first experiment, Dubey \emph{et al.} [Phys. Rev. C \textbf{112}, L011602 (2025)], we established the influence of the neutron shell closure. In the present work, the CN $^{214}$Ra was deliberately selected to investigate the dependence of the total neutron multiplicity (\textit{M$_{\mathrm{total}}$}) on the proton number (\textit{Z}), while keeping the neutron number constant at \textit{N} = 126 in the pre-actinide region. The objective of the study is two-fold : (i) to examine the effect of proton shell closure when moving away from \textit{Z} = 82, and (ii) to correlate the present results on proton shell closure with our previous finding on neutron shell closure. We have also used the previous reported data for \textit{N} = 126 isotones $^{210}$Po, $^{212}$Rn, and $^{213}$Fr to establish the validation of shell closure. A systematic increase in \textit{M$_{\mathrm{total}}$} with increasing \textit{Z} was observed from \textit{Z} = 82 to \textit{Z} = 88. Furthermore, comparison of the present results on proton shell closure with our earlier neutron shell closure observation, reveals that the cross-correlation between neutron and proton shell closure shows a systematic increase in \textit{M$_{\mathrm{total}}$} as one moves away from $^{208}$Pb, whether along isotonic or isotopic chains.
 
\end{abstract}
\maketitle
\textit{Introduction.} Heavy-ion-induced fusion-fission (HIFF) continues to be a subject of active investigation due to the significant influence of the entrance channel effects on various observables, such as the mass and angular distributions of fission fragments, total kinetic energy, and particle emission multiplicities. Many experiments have looked into how nuclear shell effects and energy dissipation influence the HIFF process, especially in reactions close to the Coulomb barrier  \cite{01,02,03,04,05,06,07,08}. One of the primary challenges in comprehending HIFF lies in addressing the phenomenon known as fission hindrance. This means that the fission process is slowed down by dissipative forces, causing a delay before the nucleus splits. This delay gives the excited CN more time to emit particles like neutrons, light charged particles, and gamma ($\gamma$) rays, which increases its chances of surviving fission for longer duration. Among these emissions, the number of neutrons released before the nucleus splits called pre-scission neutron multiplicity (\textit{M$_{\mathrm{pre}}$}) and is a useful tool to study how long the fission process takes place \cite{09}. These neutrons can be separated from those released after fission because they have different kinematic signatures, making their measurement more accurate. Interestingly, the number of these pre-scission neutrons observed in experiments is much higher than those predicted by standard statistical models \cite{09,10}. Similar high emission levels have also been observed for pre-scission light charged particles \cite{11} and Giant Dipole Resonance $\gamma$-rays  \cite{12}. \\
Neutron multiplicity has been extensively studied through numerous experiments employing a diverse array of projectile–target combinations \cite{01,02,03,04,05,06,13,14,15,16,17,18,19,20,21,22,23}, intending to elucidate the effects of various physical parameters. These investigations have focused on factors such as entrance channel characteristics, nuclear shell closures, \textit{N/Z}, and excitation energy. Additionally, nuclear dissipation has been identified as a key component influencing the HIFF and it's significant role in governing nuclear behavior has been consistently observed across different mass regions in a variety of experimental and theoretical studies \cite{02,03,04,16,17,18,19,20,24,25,26,27,28,29,30,31,32}. Various other investigations have indicated that nuclear dissipation is comparatively weaker in nuclei exhibiting shell closure than in the neighbouring nuclei \cite{03,33}. Additionally, several studies have demonstrated that the influence of shell effects on \textit{M$_{\mathrm{pre}}$} diminishes with increasing excitation energy, evolving from a dominant to a marginal contribution \cite{02,33}. To probe the role of shell closure, specific attention has been given to the neutron shell closure at \textit{N}=126, where it has been reported that \textit{M$_{\mathrm{pre}}$} attains a minimum near the closed shell at low excitation energies \cite{01,02,03}. However, our recent experimental findings \cite{04} reveal that shell effects remain significant even at higher excitation energies. Notably, the influence of shell effects becomes more significant in the post-scission neutron multiplicity (\textit{M$_{\mathrm{post}}$}) and is even more prominently reflected in the \textit{M$_{\mathrm{total}}$}.\\
In the present study, we have formed the shell-closed CN $^{214}$Ra through the reaction $^{28}$Si+$^{186}$W and calculated the \textit{M$_{\mathrm{pre}}$}, \textit{M$_{\mathrm{post}}$} and  \textit{M$_{\mathrm{total}}$}. Our results were compared with existing data in the literature for nuclei having the same neutron number (N=126) but different proton numbers. Focusing primarily on the pre-actinide region, we included comparisons with $^{210}$Po, $^{212}$Rn, and $^{213}$Fr.  In the present work, we have initially compared our findings with the dynamical model code \cite{34} and the experimental systematics \cite{35}. Furthermore, a systematic analysis was conducted using data from NuDat-3.0 \cite{36} and the microscopic shell correction values provided by Möller et al. \cite{37}. Finally, a cross-correlation between neutron and proton shell closures was performed based on experimental data available in the literature.\\

\textit{Experiment.} The experiment was performed at the National Array of Neutron Detectors facility, located at the Inter-University Accelerator Centre, New Delhi, India. A pulsed $^{28}$Si beam with a repetition rate of 250 ns, delivered by the 15UD Pelletron coupled with a LINAC booster, was directed onto a $^{186}$W target of an areal density 637-$\mu$g/cm$^2$. The target, backed by a 40-$\mu$g/cm$^2$ carbon layer, was prepared using the ultra-high vacuum evaporation technique \cite{38}. It was mounted at the center of a 100 cm diameter spherical reaction chamber. Measurements were carried out at five excitation energies 55.7, 61.8, 72.6, 80.0, and 91.3 MeV. A pictorial view of the complete experimental setup is presented in Fig. 1, while a schematic representation can be found in Ref. \cite{04}.
%--------------------------------------
\begin{figure}
\begin{center}
\includegraphics[width=7.0 cm] {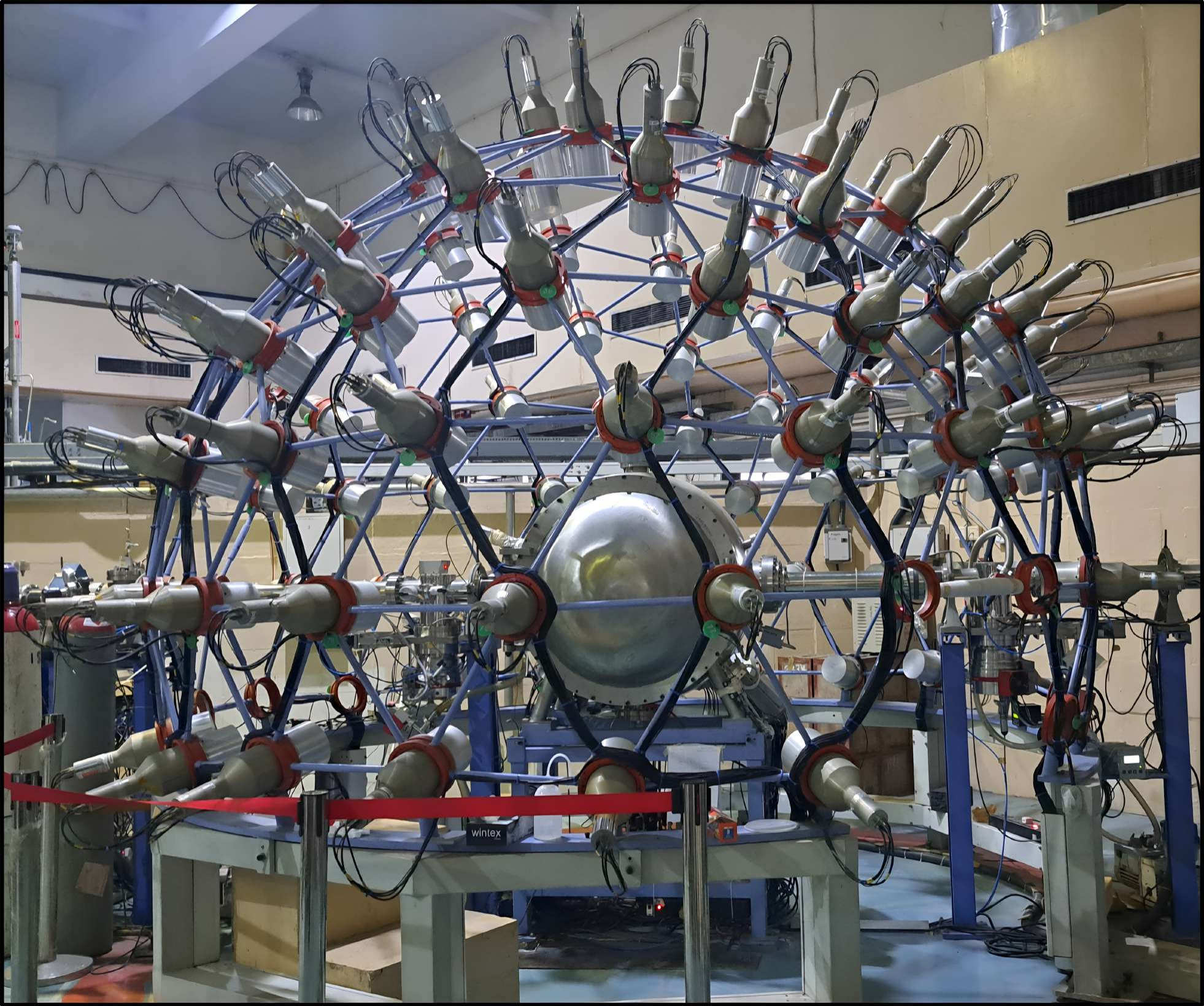}
\caption{Pictorial view of the experimental setup for measuring neutron multiplicity in coincidence with fission fragments.}
\end{center}
\end{figure} 
%---------------------%
%\begin{figure}
%\begin{center}
%\includegraphics[width=8.0 cm] {fig2.pdf}
%\caption{Time correlation spectra of complementary fission fragments detected in the MWPCs for $^{28}$Si+$^{186}$W at \textit{E}$^{*}$=80.0 MeV.}
%\end{center}
%\end{figure} 
%---------------------%
%\begin{figure}
%\begin{center}
%\includegraphics[width=8.0 cm] {PSDvsTOF_cropped.pdf}
%\caption{Pulse Shape Discrimination (PSD) vs Time of Flight (TOF) spectrum of a neutron detector for $^{28}$Si+$^{186}$W at \textit{E}$^{*}$=80.0 MeV.}
%\end{center}
%\end{figure} 
%---------------------%

%---------------------%
\begin{figure*}
\begin{center}
\includegraphics[width=16.0 cm, height=8.0 cm]{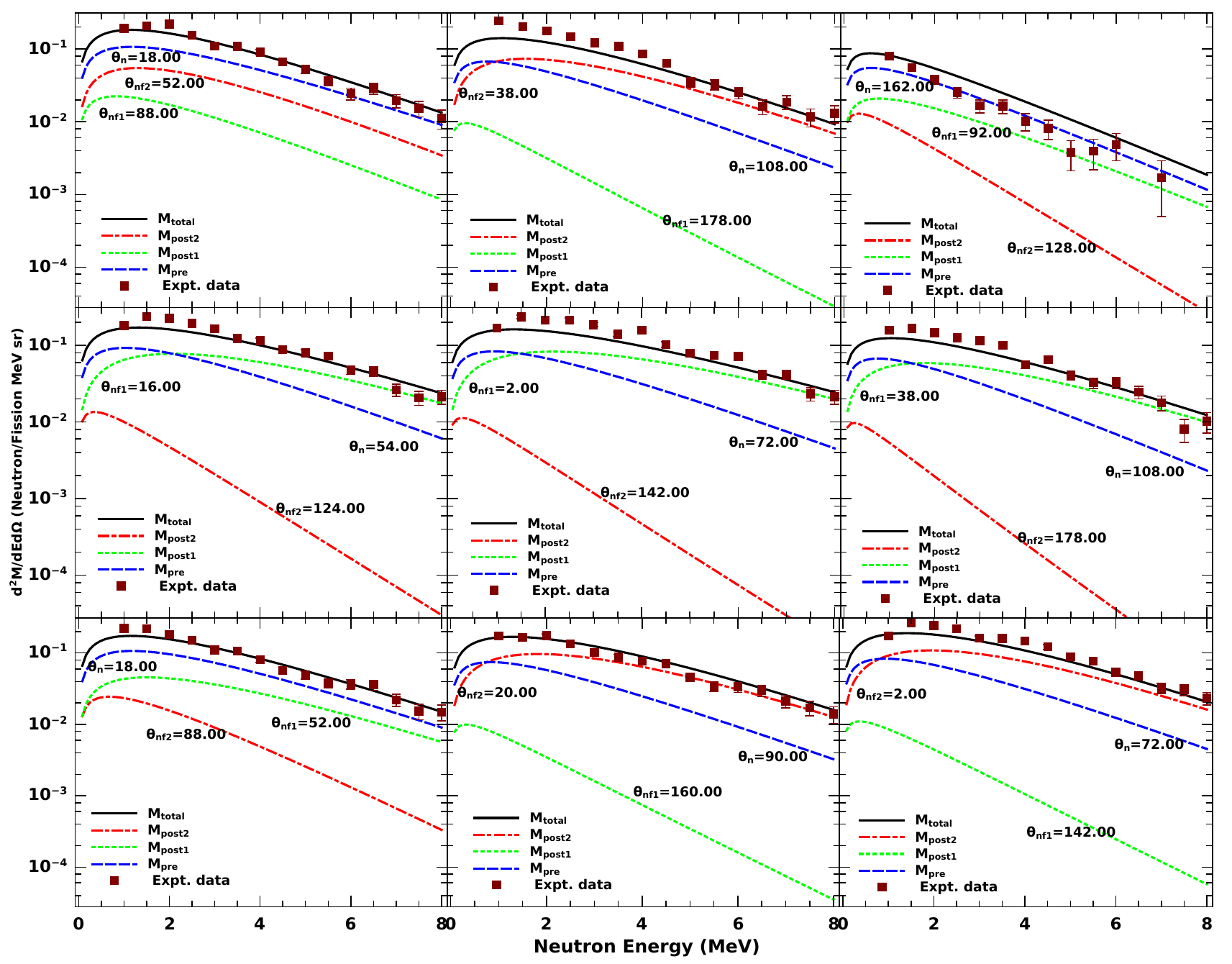}
\caption{Double differential neutron multiplicity spectra for the reaction $^{28}$Si+$^{186}$W at \textit{E$^{*}$}=80.0 MeV for neutron energies 1-8 MeV. Here, $\theta$$_{n}$ represent polar angle of neutron detector, $\theta$$_{nf_{1}}$ and $\theta$$_{nf{2}}$ are the angles between neutron detector and fission fragments. Square represents present experimental data, \textit{M$_{\mathrm{pre}}$} (dot-dot-dash), \textit{M$_{\mathrm{total}}$} (solid line),  \textit{M$_{\mathrm{post1}}$} (dotted line),  \textit{M$_{\mathrm{post2}}$} (dot-dashed) lines from the fragments.}
\end{center}
\end{figure*}
%---------------------%
Two multiwire proportional counters (MWPCs), each with an active area of 20 × 10 cm$^2$ \cite{39}, were employed to detect the complementary fission fragments (FF). These detectors were symmetrically positioned at ±70$^{\circ}$ with respect to the beam axis, at a distance of 27 cm from the target, corresponding to the folding angle between the FF. The MWPCs were operated with isobutane gas at a pressure of 4 mbar. To monitor the beam, two silicon surface barrier detectors were installed inside the chamber at $\pm$12$^{\circ}$ relative to the beam direction. Neutron detection was carried out using an array of sixteen BC501A organic liquid scintillator detectors \cite{40}. These detectors were arranged azimuthally around the reaction plane to cover an angular range from 18$^{\circ}$ to 342$^{\circ}$, with a fixed flight path of 175 cm from all the neutron detectors. The energy threshold for neutron detection was set around 0.5 MeV, calibrated using $\gamma$-ray sources ($^{137}$Cs and $^{60}$Co) \cite{41}. Background neutron contributions were minimized using a beam dump positioned 4.5 m downstream from the target, along with appropriate shielding consisting of paraffin and lead bricks. \\
\begin{table*}
\caption{\label{tab:table1}Experimentally calculated results for $^{28}$Si+$^{186}$W reaction forming $^{214}$Ra at various excitation energies.}
\begin{ruledtabular}
\begin{tabular}{cccccc}
 \textit{\textit{E$^{*}$}} (MeV) &$M_{\mathrm{pre}}$ & $M_{\mathrm{post}}$ & $M_{\mathrm{total}}$ &$T_{\mathrm{pre}}$ & $T_{\mathrm{post}}$ \\ \hline
 55.7 & 2.14$\pm$0.15 & 1.58$\pm$0.06 & 5.30$\pm$0.17 & 1.40$\pm$0.07 & 1.34$\pm$0.04  \\
 61.8 & 2.47$\pm$0.17 & 1.67$\pm$0.07 & 5.81$\pm$0.20 & 1.46$\pm$0.07 & 1.26$\pm$0.04  \\
 72.6 & 2.71$\pm$0.15 & 1.77$\pm$0.06 & 6.25$\pm$0.17 & 1.50$\pm$0.07 & 1.16$\pm$0.03  \\
 80.0 & 3.22$\pm$0.18 & 1.85$\pm$0.08 & 6.92$\pm$0.21 & 1.47$\pm$0.08 & 1.10$\pm$0.03  \\
 91.3 & 3.41$\pm$0.20 & 2.15$\pm$0.09 & 7.71$\pm$0.24 & 1.26$\pm$0.08 & 1.05$\pm$0.03  \\
\end{tabular}
\end{ruledtabular}
\end{table*}

\textit{Data Analysis and Results.} The binary fission events were analyzed using the ROOT-based data analysis \cite{42}. Fission fragments were distinguished from other charged particles, such as scattered projectiles and target recoils, through time-of-flight (TOF) measurements and kinematic coincidence techniques. More details regarding extraction of \textit{M$_{\mathrm{pre}}$}, \textit{M$_{\mathrm{post}}$}, \textit{M$_{\mathrm{total}}$}, \textit{T$_{\mathrm{pre}}$}, and \textit{T$_{\mathrm{post}}$} can be found in Ref. \cite{04} and results of calculation are given in Table 1 and presented in Fig. 2.
\begin{figure}
\begin{center}
\includegraphics[width=8.0 cm, height=10.0 cm]{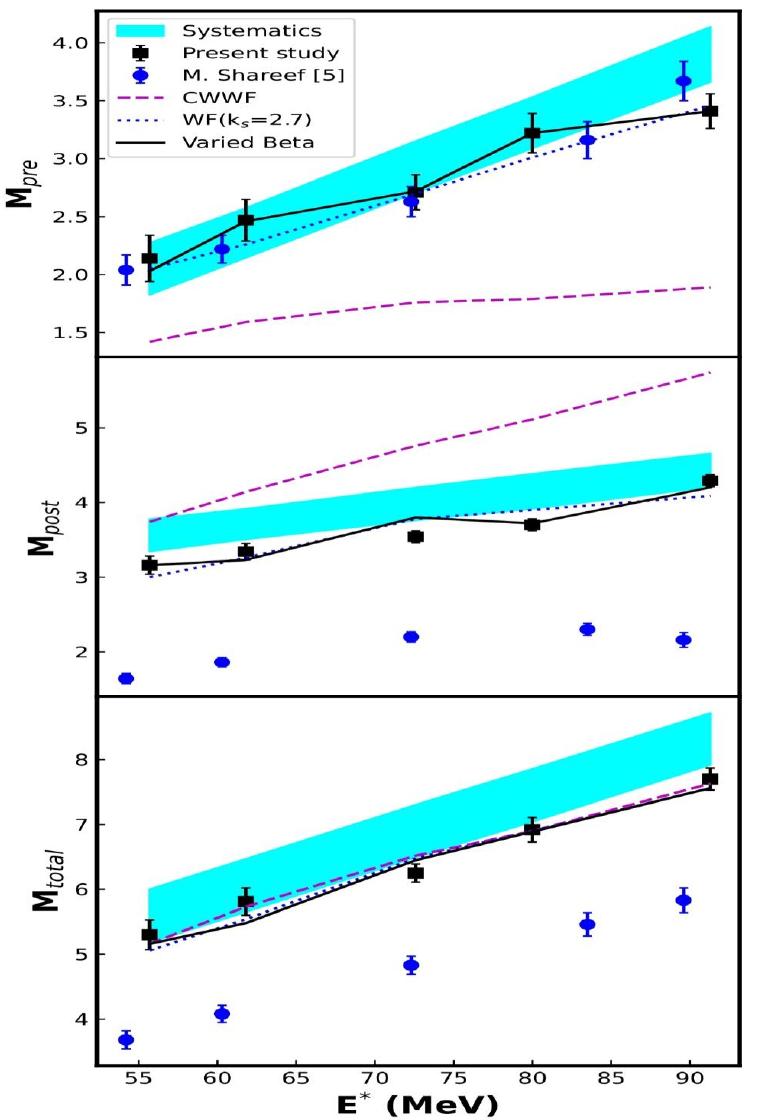}
\caption{Variation of experimentally and theoretically calculated \textit{M$_{\mathrm{pre}}$} (top), {M$_{\mathrm{post}}$} (middle) and \textit{M$_{\mathrm{total}}$} (bottom) w.r.t excitation energy for compound nucleus $^{214}$Ra.}
\end{center}
\end{figure}
%---------------------%

\textit{Discussion.} The dynamical model employs the stochastic Langevin equation to investigate the full-time evolution of an excited compound nucleus. This analysis starts from the system's initial ground-state configuration and proceeds up to the point of scission. The collective coordinate used in this model is the Funny-Hills shape parameter \textit{c} \cite{43}, which quantifies the elongation of the nuclear shape. The one-dimensional form of the Langevin equation, expressed in terms of the parameter \textit{c}, is given by \cite{44, 45, 46}:
\begin{equation}
{\frac{dp}{dt} = - \frac{p^{2}}{2} \frac{d}{dc} \left(\frac {1} {{m(c)}}\right)-\frac{dF}{dc}-{\eta(c)}p+g\Gamma(t),}
\end{equation}
\begin{equation}
\frac{dc}{dt} =  \frac{p}{m(c)},
\end{equation}
Here, \textit{p} represents the momentum conjugate to the deformation parameter \textit{c}. According to the fluctuation-dissipation theorem, the strength of the stochastic (random) force \textit{g} is related to the friction coefficient $\eta$ by the relation \textit{g} = $\sqrt{\eta T}$ \cite{47}, where \textit{T} denotes the temperature. Further details regarding the implementation of this model can be found in Refs. \cite{04, 34}. In the present study, we measured the neutrons emitted from both the fissioning nucleus and the resulting FF. For each incident beam energy, the quantities \textit{M$_{\mathrm{pre}}$}, \textit{M$_{\mathrm{post}}$} and \textit{M$_{\mathrm{total}}$} were estimated by averaging over an ensemble of 10$^{6}$ Langevin simulation events.\\
Several prescriptions for the dissipation coefficient have been employed to reproduce the experimentally observed values of \textit{M$_{\mathrm{pre}}$}, \textit{M$_{\mathrm{post}}$} and \textit{M$_{\mathrm{total}}$}. As depicted in Figure 3, conventional theoretical frameworks, the chaos-weighted wall friction (CWWF) model \cite{48,49}, and the wall-plus-window friction (WF) model when employed with reduction factors \textit{k$_s$}=0.25 \cite{50} failed to reproduce the experimental values of \textit{M$_{\mathrm{pre}}$} and \textit{M$_{\mathrm{post}}$} at the corresponding excitation energies. A significantly improved agreement with experimental data is obtained when the WF model is utilized with an enhanced reduction factor of \textit{k$_s$}=2.7. In addition, we conducted calculations using a shape-independent reduced dissipation coefficient ($\beta$), treated as an adjustable parameter. Figure 3 also demonstrates that selected values of $\beta$ specifically, 20.5, 20.0, 10.9, 16.5, and 12.6 MeV/$\hbar$, yield good agreement with the experimental \textit{M$_{\mathrm{pre}}$} and \textit{M$_{\mathrm{post}}$} values at \textit{E}$^{*}$= 55.7, 61.8, 72.6, 80.0 and 91.3 MeV. Similarly, we computed \textit{M$_{\mathrm{total}}$} using all the dissipation prescriptions discussed earlier and found that each yields comparable results, all of which show excellent agreement with our experimental measurements. To further validate the accuracy of our data, we also calculated \textit{M$_{\mathrm{pre}}$}, \textit{M$_{\mathrm{post}}$} and \textit{M$_{\mathrm{total}}$} using the empirical systematics proposed by Kozulin et al. \cite{35}. The results obtained from these calculations are consistent with the systematic trend indicated by the shaded region in Fig. 3. In this figure, we also included a comparison with the experimental data reported by Shareef \textit{et al.} \cite{05}. The comparison reveals that both datasets yield comparable values for the \textit{M$_{\mathrm{pre}}$}; however, the data from Shareef \textit{et al.} underestimates the \textit{M$_{\mathrm{post}}$}, which consequently leads to a lower value of the \textit{M$_{\mathrm{total}}$}. This discrepancy is likely attributable to quasi-fission \cite{51}.\\ 
%---------------------%
\begin{figure}
\begin{center}
\includegraphics[width=8.0 cm, height=6.0 cm]{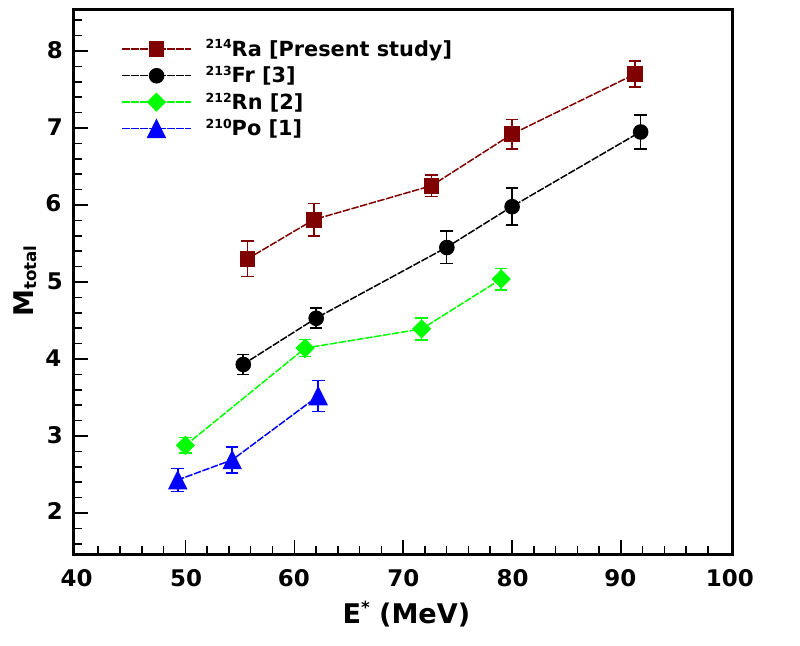}
\caption{Variation in \textit{M$_{\mathrm{total}}$} with increase in \textit{E$^{*}$} for shell closure compound nucleus $^{210}$Po, $^{212}$Rn, $^{213}$Fr and $^{214}$Ra.}
\end{center}
\end{figure}
%---------------------%
\begin{figure}
\begin{center}
\includegraphics[width=8.0 cm, height=10.3 cm]{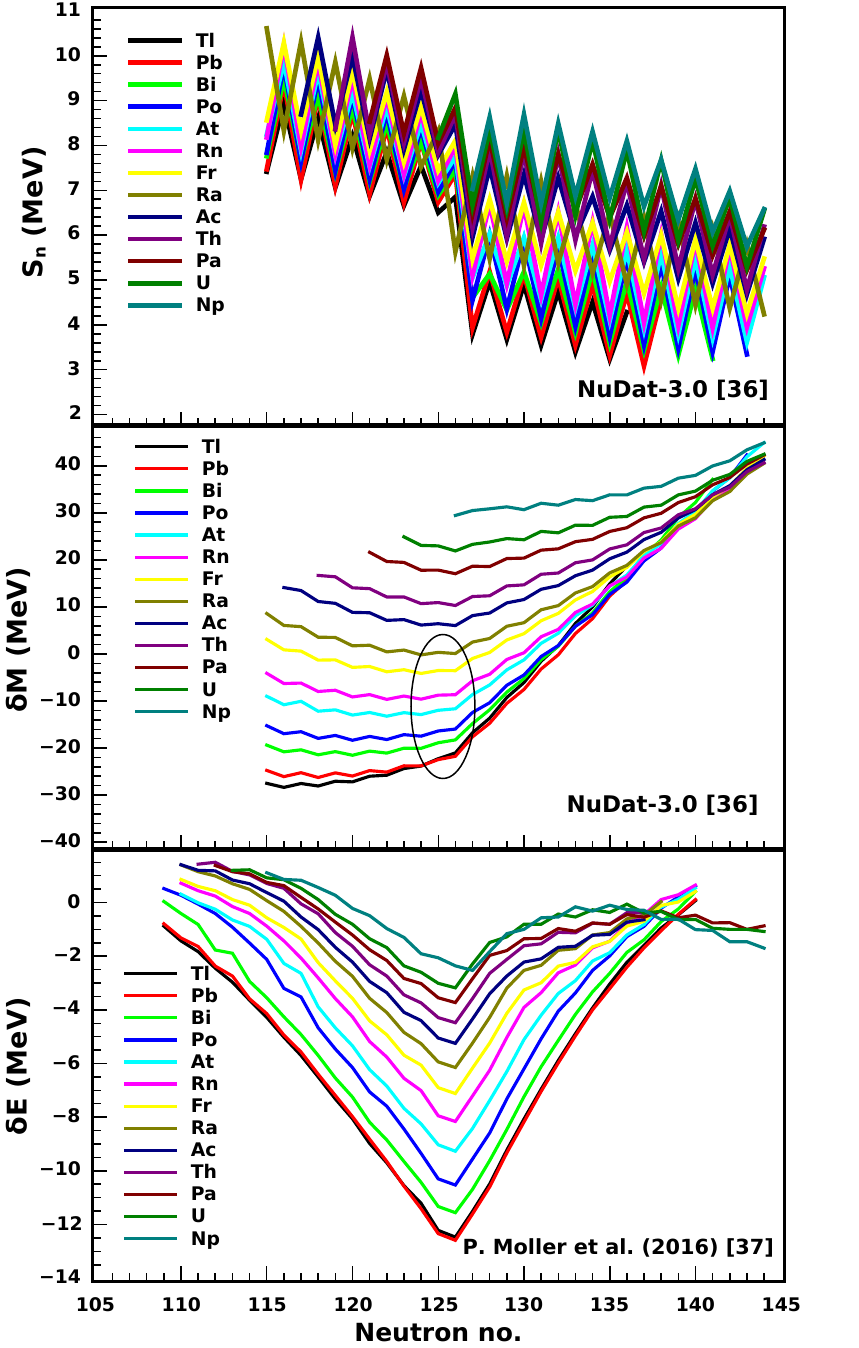}
\caption{Variation in S$_{n}$ (top), $\delta$M (middle) and $\delta$E (bottom) with increases in Neutron no. for compound nucleus Z = 81 to Z = 93.}
\end{center}
\end{figure}
 %---------------------%
 \begin{figure*}
\begin{center}
\includegraphics[width=17.2 cm, height=10.5 cm]{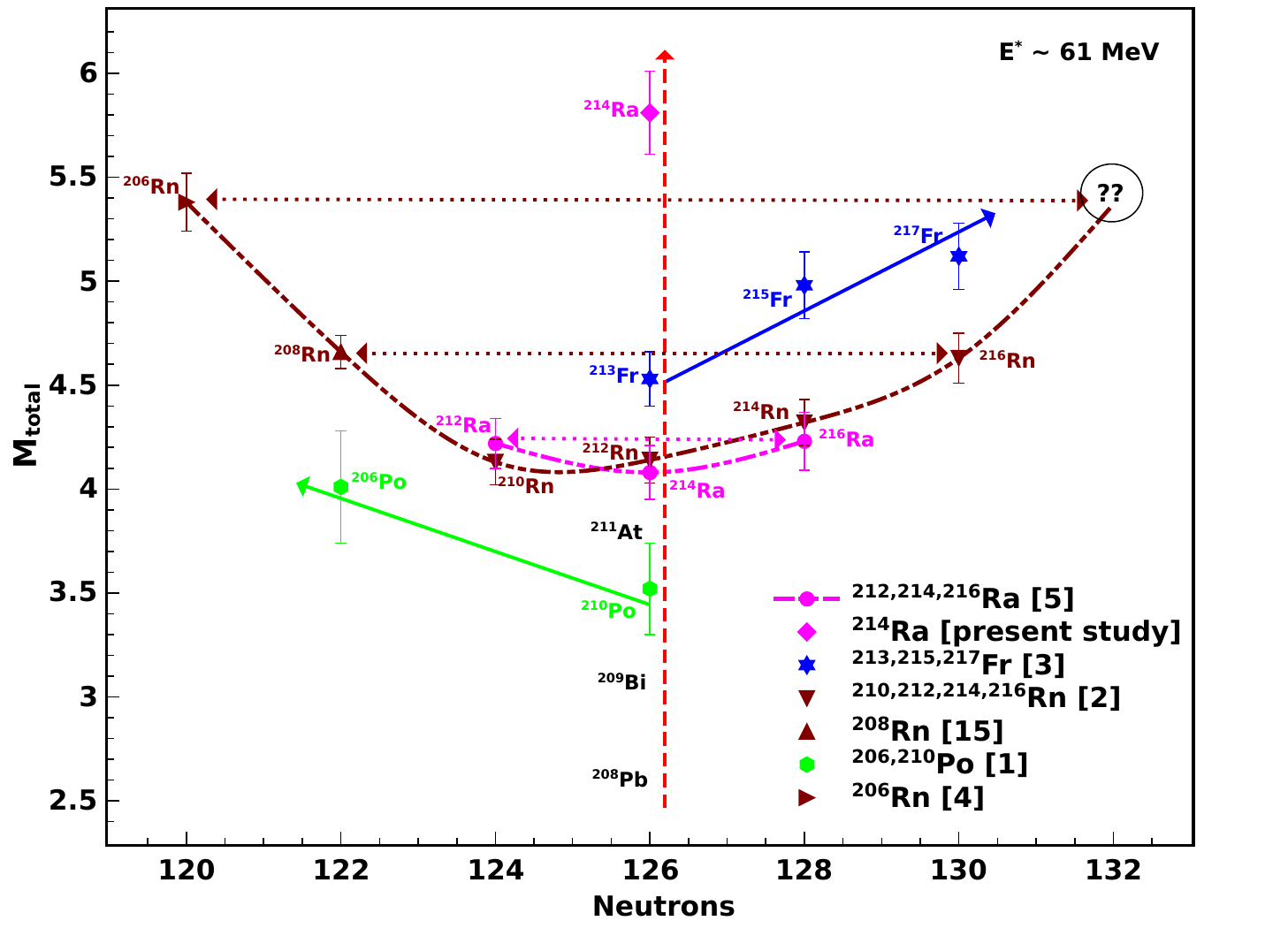}
\caption{Cross-correlation between neutron and proton shell closure with respect to \textit{M$_{\mathrm{total}}$} for pre-actinides.}
\end{center}
\end{figure*}
To investigate the influence of shell closure effects, we analyzed our current measurement of \textit{M$_{\mathrm{total}}$} for the reaction $^{28}$Si+$^{186}$W, which forms the CN $^{214}$Ra, in conjunction with existing experimental data for CN $^{210}$Po, $^{212}$Rn and $^{213}$Fr \cite{01,02,03}. In this study, the neutron number was fixed at \textit{N}=126 for all systems, while the proton number (\textit{Z}) was systematically varied within the pre-actinide region. The experimental values of \textit{M$_{\mathrm{total}}$} for these reactions at different excitation energies are presented in Fig. 4. \textit{M$_{\mathrm{total}}$} increases with increasing \textit{Z} as shown in this figure. This trend suggests that the influence of the proton shell closure at \textit{Z}=82 diminishes progressively as the system moves toward higher \textit{Z} values (up to \textit{Z}=88 ), indicating a reduced shell effect with increasing \textit{Z} from the closed proton shell.
\\In this study, our findings were further examined through a systematic analysis incorporating data from the NuDat-3.0 \cite{36}, along with microscopic shell correction values obtained from the Möller mass table \cite{37}. The primary objective was to investigate the variation of neutron separation energy (\textit{S$_{n}$}), mass excess ($\delta$M), and microscopic shell corrections ($\delta$E) as functions of changing neutron and proton numbers. To understand these trends, we analyzed data for elements with atomic numbers ranging from \textit{Z}=81 to \textit{Z}=93. The variation of \textit{S$_{n}$} (top), $\delta$M (middle), and $\delta$E (bottom) as a function of neutron number is presented in Fig. 5. As shown in the top panel of Figure 5, a pronounced drop in \textit{S$_{n}$} is observed at \textit{N}=126, indicating the presence of a strong shell closure. This drop is most prominent for lead (Pb) nuclei and becomes progressively less distinct with increasing proton number in the pre-actinide region. In the actinide region, the signature of the shell closure at \textit{N}=126 is substantially reduced. A similar trend is observed in the middle panel of Figure 5, where a sharp increase in $\delta$M is evident at \textit{N}=126 for Pb nuclei. As the proton number increases, this rise becomes progressively less pronounced, as highlighted by the elliptical loop indicated in the figure. In the bottom panel of Figure 5, a distinct minimum in the $\delta$E is observed at \textit{N}=126 across all proton numbers. This dip is most pronounced for Pb nuclei and gradually diminishes with increasing proton number. When the results from all three panels are seen collectively, it indicates that the signatures of neutron shell closure effects are strongest at \textit{Z}=82 and progressively weaken with increasing \textit{Z} in the pre-actinide region. In the actinide region, the influence of the proton shell closure is found to be minimal.
\\
Finally, we conducted a comparative analysis between our current experimental findings with our previous results reported in Ref. \cite{04}, as well as with the available literature data on  \textit{M$_{\mathrm{total}}$} \cite{01,02,03,05,15}. This comparison aims to investigate the combined influence of proton and neutron shell effects in the pre-actinide region. The outcomes of our analysis are presented in Fig. 6. In this figure, the dashed upward-pointing arrows indicate that \textit{M$_{\mathrm{total}}$} increases with increasing proton number. The solid arrows (pointing both left and right) suggest that \textit{M$_{\mathrm{total}}$} also increases as the neutron number deviates either above or below from the neutron shell closure, thereby supporting our earlier observations \cite{04}. Additionally, the double-headed arrows indicate that adding or removing an equal number of neutrons from the neutron shell closure results in similar \textit{M$_{\mathrm{total}}$} values, emphasizing the symmetry of this effect. The question mark in Fig. 6 denotes a predicted value; specifically, it is anticipated that the experimentally measured \textit{M$_{\mathrm{total}}$} for the CN $^{218}$Rn will be approximately equal to that of $^{206}$Rn. By applying a similar line of reasoning, we estimated the expected \textit{M$_{\mathrm{total}}$} values for other CN in this mass region. For instance, the positions of $^{208}$Pb, $^{209}$Bi, and $^{211}$At have been inferred and placed accordingly based on these predictive considerations. Also, the relatively lower \textit{M$_{\mathrm{total}}$} values observed for $^{212,214,216}$Ra could potentially be attributed to quasi-fission contributions \cite{05,51,52,53}.\\
\textit{Conclusion.} In summary, the effect of proton shell closure was investigated by fixing the neutron number at \textit{N}=126. Our results show that the total neutron multiplicity increases progressively from \textit{Z}=82 to \textit{Z}=88 within the pre-actinide region. This trend reflects the strong influence of the proton shell closure in Pb nuclei, which gradually weakens as one moves toward Ra compound nuclei (\textit{Z}=88). Furthermore, a cross-correlation between neutron and proton shell closures revealed that the total neutron multiplicity increases systematically as one moves away from $^{208}$Pb, either along isotonic or isotopic chains. These experimental observations corroborate theoretical predictions, demonstrating that total neutron multiplicity serves as an effective probe for identifying shell effects in heavy-ion-induced fission.\\

%---------------------%
\textit{Acknowledgements.} The authors gratefully acknowledge the Pelletron+LINAC group and the Target Laboratory at IUAC for their invaluable support during the experiment. One of the author, P. Dubey, acknowledges financial support from the Prime Minister’s Research Fellowship (PMRF) scheme. One of the author, A. Kumar extends his gratitude to the Institutions of Eminence (IoE), Banaras Hindu University [Grant No. 6031-B], and IUAC-UGC, Government of India (Sanction No. IUAC/XIII.7/UFR-71353), for their support. The authors also acknowledge the National Supercomputing Mission (NSM) for providing computational resources on PARAM Shivay at the Indian Institute of Technology (BHU), Varanasi. This facility is implemented by C-DAC and supported by the Ministry of Electronics and Information Technology (MeitY) and the Department of Science and Technology (DST), Government of India.

\bibliography{basename of .bib file}

\end{document}